# A new framework for climate sensitivity and prediction: a modelling perspective

**Francesco Ragone[1,2], Valerio Lucarini[2,3,4], Frank Lunkeit[2]**

1) Klimacampus, Institut für Meereskunde, University of Hamburg, Hamburg, Germany
2) Klimacampus, Meteorologisches Institut, University of Hamburg, Hamburg, Germany.
3) Department of Mathematics and Statistics, University of Reading, Reading, UK
4) Walker Institute for Climate System Research, University of Reading, Reading, UK

**Abstract** The sensitivity of climate models to increasing $CO_2$ concentration and the climate response at decadal time-scales are still major factors of uncertainty for the assessment of the long and short term effects of anthropogenic climate change. While the relative slow progress on these issues is partly due to the inherent inaccuracies of numerical climate models, this also hints at the need for stronger theoretical foundations to the problem of studying climate sensitivity and performing climate change predictions with numerical models. Here we demonstrate that it is possible to use Ruelle's response theory to predict the impact of an arbitrary $CO_2$ forcing scenario on the global surface temperature of a general circulation model. Response theory puts the concept of climate sensitivity on firm theoretical grounds, and addresses rigorously the problem of predictability at different time-scales. Conceptually, our results show that performing climate change experiments with general circulation models is a well-defined problem from a physical and mathematical point of view. Practically, our results show that considering one single $CO_2$ forcing scenario is enough to construct operators able to predict the response of climatic observables to any other $CO_2$ forcing scenario, without the need to perform additional numerical simulations. We also introduce a general relationship between climate sensitivity and climate response at different time scales, thus providing an explicit definition of the inertia of the system at different time scales. While what we report here refers to the linear response, the general theory allows for treating nonlinear effects as well. Our results pave the way for redesigning and interpreting climate change experiments from a radically new perspective.

**Corresponding author**

Francesco Ragone

Institut für Meereskunde, University of Hamburg

Bundesstrasse 53, D-20146 Hamburg, Germany

Tel: +49 (0)40 42838 5729

Email: francesco.ragone@zmaw.de

# 1 Introduction

One of the main goals of climate science is to predict how modulations on different time scales of internal or external parameters impact the statistical properties of the climate system. Due to the observational evidence of the ongoing anthropogenic climate change, the problem of the response of the system to increasing greenhouse gases (GHGs) and in particular $CO_2$ concentration ($[CO_2]$) is of particular social and environmental relevance (IPCC 2007a, 2013). The assessment of the future impacts of climate change under a variety of $CO_2$ forcing scenarios (IPCC 2007b, 2013) and the evaluation of the climate sensitivity to the increase of $[CO_2]$ (Knutti and Hegerl 2008, Galloway et al. 2013, Otto et al. 2013) mostly rely on the use of general circulation models (GCMs).

The intense efforts put on model development by the scientific community in the past decades have led to impressive improvements in the GCMs complexity and computational performances. However, even the evaluation of the most basic measure of climate sensitivity, the Equilibrium Climate Sensitivity (ECS, the change of the globally averaged surface temperature for doubling $[CO_2]$), is still subject to large uncertaineties (Otto et al. 2013, Sherwood et al. 2014). The relationship between the ECS and the Transient Climate Response (TCR), defined as the change in the globally averaged surface temperature at the end of 1% per year increase of CO2 until its doubling (Otto et al. 2013), is also unclear. More in general, the response of the surface temperature (as well as of other relevant observables) at different time scales for $CO_2$ forcings with non-trivial temporal evolution is a poorly understood problem, emphasized by the presence of substantial uncertainties in the predictive skills of the models at decadal time scales (IPCC 2013).

These issues are certainly partially due to technical difficulties related to the inaccuracies of the GCMs. However, the very efforts put in model development in the past decades and the results achieved show that these issues can not be approached only as a technological problem to be solved by building (arguably endlessly) bigger and more complex numerical models. The need for substantial advances in the scientific ideas at the basis of climate modelling strategies (and not only in the technological tools in use) is more and more clear in the climate community (Shukla et al 2009). An illuminating perspective on mathematical frameworks suited for a theory of climate sensitivity has been recently presented (Chekroun et al. 2011a). Here we follow a complementary approach to define a robust theoretical framework for the use of GCMs in addressing the problem of climate response, sensitivity and prediction, based upon Lucarini and Sarno (2011).

In the past, several attempts have been aimed at constructing some sort of response operator for the CS able to relate a forcing (e.g. change in $[CO_2]$) with a given spatial and temporal pattern to the time dependent climate response as measured by the change of chosen climatic observables (e.g. global surface temperature). In equilibrium statistical mechanics, the fluctuation-dissipation theorem (FDT) relates the response of a system to external perturbations to its internal fluctuations in the unperturbed state (Kubo 1966). The FDT has proved to be a very powerful tool for studying a variety of physical processes in condensed matter physics, acoustics, optics and many other fields of physics. FDT thus provides a tempting framework for deducing climate response from its internal variability. In fact, varous authors have gone in this direction by adopting simplified versions of the FDT, with in general a satisfactory degree of success (Langen and Alexeev 2005, Gritsun and Branstator 2007, Abramov and Majda 2008). However, it has been shown that the applicability and effectiveness of FDT-based methods to nonequilibrium systems crucially depend on the observable one wants to predict (Cooper and Haynes 2011, Cooper et al. 2013). See also the discussion in Wouters ans Lucarini (2013). Here we approach the problem from a related but different point of view.

Ruelle introduced general methods for studying how nonequilibrium systems respond to external perturbations (Ruelle 1998a, 1998b, 2009). Ruelle's response theory (RRT) clarifies the limits of FDT-based methods for nonequilibrium systems and shows that is indeed possible to compute deviations from a nonequilibrium steady state (NESS) due to weak forcings through explicit response formulas and using the statistical properties of the unperturbed state only. Recently RRT has been proposed as a rigorous framework for computing climate response and its applicability has been tested on the Lorenz 96 model (Lucarini and Sarno 2011). Going in the direction of bridging the gap between statistical mechanical theories and climate science, we show how, starting from an ensemble of runs forced by instantaneous $CO_2$ doubling, the formalism of RRT can be used for predicting the response of the globally averaged surface temperature of a GCM to an arbitrary temporal evolution of the $CO_2$ forcing. Despite the nonlinear nature of the system, the statistical properties of the response are remarkably well captured by the linear version of RRT even when finite forcings of practical interest are applied. Our results show that in order to predict the response of the system to any $CO_2$ forcing scenario at both infinite and lead-time, only one single scenario needs to be taken into account. This suggest that one could reduce the vast range of forcing scenarios considered in the IPCC protocol with few selected scenarios, and derive more information from currently available data. While what we report here refers to linear response, the theory behind these findings allows for treating nonlinear effects as well (Lucarini 2009, Lucarini and Colangeli 2012).

Note that this approach targets the way we use numerical models in climate science. The evaluation of the sensitivity of the real climate depends of course on the quality and realism of the models we use. In this sense, RRT does not lead directly to an improvement of the skills of the models in representing the actual climate and climate change. What RRT inform us about is not the realism of the answer, but the consistency of the following scientific question: how does the state of a GCM change when changing one or more of its parameters? If this question proved to be ill posed (that in general with a nonequilibrium system like a complex GCM could be), the whole project of the numerical simulation of climate change would be essentially flawed, independently on the (apparent) level of realism of the results. The applicability itself of RRT demonstrates that the problem of studying climate change with GCMs is mathematically and physically well-posed, and faceable with rigourous techniques widely used in other fields of physics. RRT provides a way to reformulate the question of climate change numerical simulation in a more theoretically grounded way, which hopefully would also help in understanding how to improve the quality of the answer and of the tools used in order to get to it. The issue of understanding and improving the skills of climate models per se is a challenging one, but of essentially different nature (IPCC 2007a, 2013).

The paper is organized as follows. In Section 2 we give a brief summary of the basic concepts of Ruelle's linear response theory, we describe the numerical experiments we have performed, and we describe how we have analyzed the data. In Section 3 we present the results of the numerical simulations and data analysis demonstrating the predictive power of the theory, we show how the concept of climate sensitivity assumes a natural role in the context of linear response theory, and we present some guidelines for designing future experiments based on a scale analysis. Eventually, in Section 4 we discuss our results and we propose possible future lines of research.

**2 Methods and materials**

*2.1 Linear response theory*

RRT strictly applies to Axiom A dynamical systems (Ruelle 1998a, 1998b, 2009). Roughly speaking, a dynamical system is Axiom A if 1) the dynamics is uniformly hyperbolic on the attractor and 2) the set of the periodic points is dense on the attractor. Axiom A systems possess a Sinai-Ruelle-Bowen (SRB) invariant measure, which guarantees a) the asymptotic equivalence of time and ensemble averages of observables (that it is not, despite intuition, a general property of nonequilibrium systems) and b) the stability of the statistical properties when a weak stochastic forcing is applied. In this sense Axiom A systems are good physical systems, for which the statistical properties of the observables are well defined. Note that property b) requires to be far from tipping points, in the neighbourhood of which response formulas are expected to fail (indeed a tipping point could be rigorously defined by the breaking of the applicability of RRT).

Demonstrating that a dynamical system is Axiom A is a difficult task even for simple systems, and impossible for a numerical model of the complexity of a GCM. However, the use of response formulas in most cases of physical interest is justified thanks to the Chaotic Hypothesis (Gallavotti 1996), which states that chaotic systems with many degrees of freedom effectively behave as Axiom A systems in terms of properties a) and b) even if they do not satisfy rigorously requirements 1) and 2), at least when considering the statistical properties of coarse-grained observables (e.g. globally or regionally integrated quantities). See also Penland (2003) for a discussion in a geophysical context. In a sense the Chaotic Hypothesis is the extension to nonequilibrium systems of the ergodic hypothesis for equilibrium systems. As an example, when we compute the expectation value of an observable in a numerical model as the long-term average on a stationary state, we are in fact implictly assuming that the system is Axiom A-like. Similarly, operations like the computation of Lyapunov exponents or the setup of data assimilation systems (Kalnay 2003) implicitly assume the same conditions. In this paper we assume that the Chaotic Hypotesis allows using RRT to study climate response with a GCM. Note that in this sense the high dimensionality and complexity of the system enforce the applicability of theoretical instruments, rather than the opposite. If we consider low dimensional models, or dynamical systems living at the edge of chaos, the basic tenets of response theory may fail (Chekroun et al. 2011b).

Let us briefly recapitulate the basic elements of RRT. See Ruelle (2009) for a detailed presentation and Lucarini and Sarno (2011) for a geophysical oriented approach. Let us consider the dynamical system described by the set of ordinary differential equations corresponding to a numerical model of the climate system. We apply a weak forcing to the system so that the evolution equations can be written as $\dot{\boldsymbol{x}} = \boldsymbol{F}(\boldsymbol{x}) + \boldsymbol{X}(\boldsymbol{x})f(t)$, where $\boldsymbol{x}$ is the state vector of the system, $\boldsymbol{F}(\boldsymbol{x})$ represents the unperturbed dynamics with set boundary conditions, $\boldsymbol{X}(\boldsymbol{x})$ is a vector field defining the pattern of the forcing in the phase space, and $f(t)$ is the time modulation of the forcing. We consider the case where the base dynamics is autonomous (no explicit time dependence in the unperturbed dynamics, e.g. absence of daily or seasonal cycle), but one could accommodate more general time-dependent flows by resorting to the concept of pullback attractor (Chekroun et al. 2011a). Ruelle showed that the expectation value of an observable $\Phi$ in the forced system can be computed as a perturbative expansion

$$\langle\Phi\rangle_f(t) = \langle\Phi\rangle_0 + \sum_{n=1}^{+\infty}\langle\Phi\rangle_f^{(n)}(t) \tag{1}$$

where $\langle\Phi\rangle_0$ is the expectation value in the unperturbed state and the perturbative terms $\langle\Phi\rangle_f^{(n)}(t)$ can be computed as convolution integrals of $f(t)$ and suitable functions determined by the unperturbed dynamics only (Ruelle 1998a, 1998b, 2009). The linear response is given by the first term in the series, that can be computed as

$$\langle\Phi\rangle_f^{(1)}(t) = \int_{-\infty}^{+\infty} d\sigma_1 G_{\boldsymbol{\Phi}}^{(1)}(\sigma_1) f(t-\sigma_1) \tag{2}$$

where $G_{\boldsymbol{\Phi}}^{(1)}(t)$ is the first order Green function of the observable $\Phi$, that can be computed with an explicit formula defined by the unperturbed flow only as described in Ruelle (1998a, 1998b, 2009).

Response theory guarantees general properties for $G_{\boldsymbol{\Phi}}^{(1)}(t)$ and its Fourier transform $\chi_{\boldsymbol{\Phi}}^{(1)}(\omega)$, the (linear) susceptibility of the observable Φ (Lucarini et al. 2005, Ruelle 2009). The Green function is in general a causal function, that is $G_\Phi^{(1)}(t) = 0$ if $t < 0$. By taking the Fourier transform of Equation (1) we have

$$\widetilde{\langle\Phi\rangle}_f^{(1)}(\omega) = \chi_\Phi^{(1)}(\omega)\tilde{f}(\omega) \qquad (3)$$

where $\tilde{f}(\omega)$ is the Fourier transform of $f(t)$. The susceptibility gives therefore the structure of the response of the system to forcings at different frequencies (time-scales). The real and imaginary parts of the $\chi_\Phi^{(1)}(\omega)$ represent the in- and out-of-phase response of the system respectively to a sinusoidal forcing at frequency $\omega$. Maxima in the absolute value of the susceptibility correspond to resonances of the system, where the response is enhanced due to positive feedbacks at the corresponding time-scales. Being $G_\Phi^{(1)}(t)$ a causal function, its Fourier transform $\chi_\Phi^{(1)}(\omega)$ obeys the following identity (Lucarini et al. 2005):

$$\chi_\Phi^{(1)}(\omega) = \frac{i}{\pi} P \int_{-\infty}^{\infty} d\omega' \frac{\chi_\Phi^{(1)}(\omega')}{\omega' - \omega} \qquad (4)$$

where P indicates integration in principal part and the susceptibility is related to its complex conjugate through $\chi_\Phi^{(1)}(\omega) = [\chi_\Phi^{(1)}(-\omega)]^*$. Equation (4) can be recast in terms of the so called Kramers-Kronig relations (KK), self-consistency relations linking the real and imaginary parts of $\chi_\Phi^{(1)}(\omega)$ of wide use in several fields of physics, notably in optics (Lucarini et al. 2005, Lucarini 2008). RRT provides similar formulas to compute any order of nonlinearity of the response and the related nonlinear Green functions and susceptibilities (Lucarini 2009, Lucarini and Colangeli 2012). Note that limiting the attention to the linear response does not imply neglecting the nonlinear nature of the dynamics: the linear susceptibility accounts for the linear part of the response of the full nonlinear system.

Equations (2-3) show that RRT boils down to a surprisingly simple formalism for computing the linear response of a system to an arbitrary forcing, once the Green function and the susceptibility are computed. When dealing with an high dimensional system like a GCM, the direct computation of $G_\Phi^{(1)}(t)$ on the unperturbed dynamics following Ruelle (1998a,1998b,2009) can be extremely difficult. However, the above formulas allow computing $G_\Phi^{(1)}(t)$ and $\chi_\Phi^{(1)}(\omega)$ from a set of suitabe forcing experiments in a simple way. Consider performing an experiment with a prescribed forcing $f(t)$. Let us suppose that we are in the linear regime, so that the observed response of the observable Φ is approximately $\langle\Phi\rangle_f^{(1)}(t)$. Equations (2-3) provide then the basis for deriving $G_\Phi^{(1)}(t)$, or equivalently $\chi_\Phi^{(1)}(\omega)$, from the output of the numerical experiments and the (known) functional form of the forcing (Lucarini and Sarno 2011, Lucarini et al. 2014). Note that Equation (3) implies that for reconstructing $\chi_\Phi^{(1)}(\omega)$ one must consider a broadband forcing, since the response $\langle\widetilde{\Phi}\rangle_f^{(1)}(\omega)$ does not contain frequencies that are not present in the forcing $\tilde{f}(\omega)$.

Knowing $G_\Phi^{(1)}(t)$, we can then use Equation (2) to perform projections at any lead-time *t* when a different temporal evolution $g(t)$ of the forcing is used, while knowing $\chi_\Phi^{(1)}(\omega)$ inform us about the properties of the response of the system to forcings on a wide horizon of time-scales. Demonstrations of the validity of RRT for simple systems of geophysical interest have been proposed in the past (Abramov and Majda 2008, Lucarini and Sarno 2011); here we assess the predictive power of the theory with a real GCM, following the approach described above.

*2.2 Experimental settings*

*2.2.1 Description of the model*

The numerical model used in this study is PLASIM (Fraedrich et al. 2005), a simplified GCM developed at the University of Hamburg. Even if indeed not competitive with state-of-the-art GCMs, PLASIM produces a fairly realistic present climate and is representative of the class of complex numerical models used for operational climate prediction. The dynamical core is based on the Portable University Model of the Atmosphere PUMA (Fraedrich et al. 1998). The primitive equations are solved by a spectral transform method. The model has a full set of physical parameterizations for unresolved processes. Parameterizations include long and shortwave radiation with interactive clouds, horizontal and vertical diffusion, boundary layer fluxes of latent and sensible heat. Stratiform precipitation is generated in supersaturated states, and the Kuo scheme is used for deep moist convection, while shallow cumulus convection is parameterized by means of vertical diffusion. See the Reference Manual freely available toghether with the code at http://www.mi.uni-hamburg.de/plasim for a detailed and referenced description of the parameterizations included in the model.

The atmospheric model is coupled to a 1-layer slab model of the oceanic mixed layer with a depth of 50 m, which includes a thermodynamic sea-ice module. The climate response and sensitivities evaluated with PLASIM are therefore due only to the fast feedbacks (Lunt et al. 2010, Previdi et al. 2013), missing contributions from ocean, continental ice-sheets, vegetation and interactive carbon cycle. For the present study we set the model at T21 horizontal resolution and 10 levels vertical resolution, for a total of $O(10^5)$ degrees of freedom, with a time-step of 45 minutes. In order to simplify the analysis, we have removed daily and seasonal cycles, so that the evolution equations in the reference state do not explicitly depend on time.

*2.2.2 Description of the experiments*

We consider as our observable the globally averaged surface temperature $T_s$ and as forcing the convergence of radiative fluxes due to the increase in the logarithm of [$CO_2$], since the radiative forcing scales approximately logarithmically with [$CO_2$] within a reasonable range of concentrations (IPCC 2007b, 2013). Therefore from now on $\langle \Phi \rangle_f^{(1)} = \langle T_s \rangle_f^{(1)}$ represents the expectation value of the increase the globally averaged surface temperature $T_s$ in the linear regime and $f(t)$ represents the temporal evolution of the radiative forcing correspondent to a chosen $CO_2$ forcing scenario. Linear RRT in principle applies only in the limit of infinitesimal forcings. Nonetheless, we show that its range of validity extends to rather intense finite forcings.

We consider a control run of 2400 years. The [$CO_2$] is set to the value of 360 ppm, representative of the present-day value. We then perform two sets of forcing experiments, prototypical of the scenarios proposed by the IPCC protocol for estimating the climate sensitivity and the impact of CO2 forcings on the CS. In the first set of experiments we double istantaneously the CO2 concentration and we keep it fixed at 720 ppm afterwards. This corresponds to a constant radiative forcing after the [$CO_2$] doubling. This is the scenario used for computing the *ECS*. In the second set of experiments [$CO_2$] is increased by 1% per year until reaching 720 ppm (after about 70 years), and it is kept fixed afterwards. This corresponds to a radiative forcing linearly increasing for the first 70 years until [$CO_2$] has doubled, and constant afterwards. This is the scenario in which one computes the TCR. In both experiments the forcing is applied homogeneously at each point of the model.

In both cases we have performed an ensemble of 200 experiments starting from different initial conditions (Lucarini 2009, Lucarini and Sarno 2011). The 200 initial conditions are taken from the control run at intervals of 10 years starting from year 200, in order to guarantee their statistical independence and performing in this way a reasonable sampling of the attractor of the unperturbed system. Each run is 200 years long. For each set of forcing experiments we consider yearly averaged values of $T_s$ as output. We then compute the expectation values of the temperature difference $\langle T_s \rangle_f^{(1)}$ and $\langle T_s \rangle_{g_\tau}^{(1)}$ by averaging over all the ensemble members and subtracting the average value computed from the control run.

*2.2.3 Data analysis*

In general, the susceptibility $\chi_{T_s}^{(1)}(\omega)$ can be computed from the response of the surface temperature $\langle T_s \rangle_f^{(1)}(t)$ by taking its Fourier transform $\langle \widetilde{T_s} \rangle_f^{(1)}(\omega)$ and using Equation (2), being the Fourier transform of the forcing $\tilde{f}(\omega)$ known. The Green function $G_{T_s}^{(1)}(t)$ can then be derived by taking the inverse Fourier transform of $\chi_{T_s}^{(1)}(\omega)$. For special choices of the forcing, $G_{T_s}^{(1)}(t)$ can be computed directly from $\langle T_s \rangle_f^{(1)}(t)$. If we consider the case of instantaneous [CO$_2$] doubling, the time evolution of the forcing is given by $f(t) = f_{CO_2}^{2x} H(t)$, where $f_{CO_2}^{2x}$ is a constant and $H(t)$ is the Heaviside function ($H(t) = 0$ for $t \leq 0$ and $H(t) = 1$ for $t > 0$), setting $t = 0$ at the instant at which the [CO$_2$] doubles. Inserting this expression of $f(t)$ in Equation (2) and differentiating both sides of the equation gives

$$f_{CO_2}^{2x} G_{T_S}^{(1)}(t) = \frac{d}{dt} \langle T_s \rangle_f^{(1)}(t) \tag{5}$$

The Green function can therefore also be computed differentiating the response signal, and the susceptibility can then be computed by taking its Fourier transform.

All the computations have been performed in MATLAB® V. 7.9 environment using the standard Fast Fourier Tranform (fft) algorithm. When applying fft, we implictly force the input function to be periodic outside the time domain where it is defined. Applying this method to the time series of $\langle T_s \rangle_f^{(1)}$ introduces a frequency-dependent bias in the estimate of the spectrum due to the long-term behaviour of $\langle T_s \rangle_f^{(1)}$. We need to keep the information on such long-term behaviour, in order to estimate properly the low-frequency variability. It is possible to cure this issue by subtracting a suitably defined function before applying fft (Nicolson 1973).

The analysis of the spectral properties of the response has been carried out starting from Eq. (4). In this case, we have used a slightly modified version (Lucarini et al. 2005) of the KK relations usually referred to as singly subtractive Kramers-Kronig relations (SSKK). Conventional KK relations stems from considering the real and imaginary parts of Equation (4) and exploiting the symmetries of the susceptibility, obtaining

$$\Re\{\chi_{T_S}^{(1)}(\omega)\} = \frac{2}{\pi} P \int_0^{+\infty} d\omega' \frac{\omega' \Im\{\chi_{T_S}^{(1)}(\omega)\}}{\omega'^2 - \omega^2}$$

$$\Im\{\chi_{T_S}^{(1)}(\omega)\} = -\frac{2\omega}{\pi} P \int_0^{+\infty} d\omega' \frac{\Re\{\chi_{T_S}^{(1)}(\omega)\}}{\omega'^2 - \omega^2} \tag{6}$$

KK relations provide a self-consistency test for measured or observed data in many fields of physics, notably in optics (Lucarini et al. 2005). The accuracy of the KK inversion depends on the quality of the observed data. The SSKK method consists in referring to an anchor point $\omega_1$ where it is supposed that the direct estimate of $\chi_{T_S}^{(1)}(\omega)$ from the data is reliable (Lucarini et al. 2005). The SSKK relations then read

$$\Re\{\chi_{T_s}^{(1)}(\omega)\} - \Re\{\chi_{T_s}^{(1)}(\omega_1)\} = \frac{2(\omega^2 - \omega_1^2)}{\pi} P \int_0^{+\infty} d\omega' \frac{\omega' \Im\{\chi_{T_s}^{(1)}(\omega)\}}{(\omega'^2 - \omega^2)(\omega'^2 - \omega_1^2)}$$

$$\omega^{-1} \Im\{\chi_{T_s}^{(1)}(\omega)\} - \omega_1^{-1} \Im\{\chi_{T_s}^{(1)}(\omega_1)\} = -\frac{2(\omega^2 - \omega_1^2)}{\pi} P \int_0^{+\infty} d\omega' \frac{\Re\{\chi_{T_s}^{(1)}(\omega)\}}{(\omega'^2 - \omega^2)(\omega'^2 - \omega_1^2)}$$
(7)

With a careful choice of the anchor point the SSKK relations improve the accuracy of the KK analysis (Lucarini et al. 2005). In our case we have taken $\omega_1 = 2\pi\xi_1$ with $\xi_1 = 0.1\ years^{-1}$, obtaining a rather good quality of the inversion. We remark anyway that the results are rather robust with respect to different choices of the anchor point $\omega_1$.

## 3 Results

*3.1 Climate response and prediction*

Figure 1 shows the time serie of the ensemble average of the increase of globally averaged surface temperature for the instantaneous doubling scenario (blue). The shaded area represents the 95% (two standard deviations) of the ensemble variability. The long-term increase of the surface temperature for the doubling scenario (the equilibrium climate sensitivity) is rather high if compared with what is typically obtained with standard IPCC models, being 8.1 K against typical estimates between 1.5 and 4.5 K (IPCC 2007b, 2013). This is due to the fact we have chosen a simplified setup without the daily and in particular the seasonal cycle, which greatly enhances the system response to the radiative forcing. While this is mathematically more convenient at the price of decreasing the realism of our experiments, it also pushes the RRT more to its limits, thus providing a more stringent test for our approach. We have performed additional experiments increasing istantaneously the [CO2] by a factor $\sqrt{2}$ (about 510 ppm), checking that the observable indeed scales reasonably linearly with the logarithm of the [CO2], the actual behavior being only slightly sublinear (not shown).

In the insert of Figure 1 we show $G_{T_s}^{(1)}(t)$, derived from the doubling experiment output. The Green function $G_{T_s}^{(1)}(t)$ is to a first impression consistent with a relaxation process with a characteristic time scale of about 10 years, even if relevant differences emerge, as explained later. Hasselmann et al. (1993) introduced the concept of Green function of the global surface temperature of a GCM when addressing empirically the problem of the so called cold start of climate simulations. This is the warming negative bias obtained when, due to limitations in computational resources, a climate change experiment was not started from the preindustrial stationary state but from a stationary state corresponding to an higher GHGs concentration. In that case, the response of $T_s$ was fitted with a prescribed functional form and then $G_{T_s}^{(1)}(t)$ computed analitically with Equation (5). Here, instead, we are computing $G_{T_s}^{(1)}(t)$ numerically directly from the output of the GCM without loss of information.

In order to test the predictive power of the theory, we consider another scenario where [CO₂] is increased by 1% per year starting from the present-day value of 360 ppm until the value of 720 ppm is reached, and kept fixed afterwards. Since to a very good approximation the radiative absorption is proportional to the logarithm of [CO2], the corresponding radiative forcing is, to a good approximation, a ramp function $g_\tau(t) = f_{CO_2}^{2x} t/\tau$ for $0 \leq t \leq \tau$ and $g_\tau(t) = f_{CO_2}^{2x}$ for $t > \tau$ with $\tau \approx 70\ years$. In Figure 3 we compare the ensemble average of the simulations and the prediction for $\langle T_s \rangle_{g_\tau}^{(1)}$ obtained using the estimate of the Green function shown in Figure 2. The agreement is excellent both on the short and long term. Discrepancies of less that 10% are present during the transient in the window between 25 and 100 years, because of of the strong nonlinearities due to the activation of the ice-albedo feedback. The degree of precision of the prediction obtained with the linear RRT is remarkable, considering the complexity of the climate model in use and the presence of strong nonlinearities in the underlying equations.

*3.2 Climate sensitivity in response theory framework*

Figure 3 shows the real $\Re\{\chi_{T_s}^{(1)}(\omega)\}$ and imaginary $\Im\{\chi_{T_s}^{(1)}(\omega)\}$ part of the susceptibility, computed as the Fourier transform of $G_{T_s}^{(1)}(t)$. The function is plotted against the frequency $\xi = \omega/2\pi$ measured in $years^{-1}$. The susceptibility to CO₂ forcing at a frequency $\xi'$ gives a precise quantification of the climate response at the corresponding time scale $t' = 1/\xi'$. The estimate of the susceptibility becomes rather noisy for frequencies higher than $0.25\ years^{-1}$ (time-scales $\leq 4\ years$,), where we are not able to compute reliable climate projections. The overall quality of the experimentally obtained $\chi_{T_s}^{(1)}(\omega)$ is confirmed by the good agreement between the measured and KK-reconstructed real and imaginary parts. We have truncated the integral in Equation (4) at the highest measured frequency $\xi_h = 0.5/\Delta t$, where $\Delta t = 1\ year$, and used the SSKK technique for the reconstruction (Lucarini et al. 2005). One must emphasize that all the bumps around $\xi \approx 0.15\ years^{-1}$ and $\xi \approx 0.20\ years^{-1}$ found in the $\chi_{T_s}^{(1)}(\omega)$ are not due (only) to the presence of the noise, but correspond indeed to physical processes of the system (multiannual variability), as they are consistently captured by the KK relations. These spectral features are not consistent with a simple relaxation model of the climate response.

While the importance of $G_{T_s}^{(1)}(t)$ is due to its predictive power, $\chi_{T_s}^{(1)}(\omega)$ is a powerful tool for defining accurately climate sensitivity at different time scales. The equilibrium climate sensitivity is defined as $ECS = \lim_{t \to +\infty} \langle T_s \rangle^{(1)}(t)$ after an istantaneous doubling of [CO₂]. Using Equation (3) and the definition of $\chi_{T_s}^{(1)}(\omega)$ one can show that, in linear approximation,

$$ECS = f_{CO_2}^{2x}\chi_{T_s}^{(1)}(0) \tag{8}$$

Therefore *ECS* is proportional to the zero frequency value of the susceptibility. From Figure 3 (and Figure 1) we see that $f_{CO_2}^{2x}\chi_{T_s}^{(1)}(0) \approx 8.1$ K. Therefore, when computing the *ECS* as long-term average of the $T_s$ increase from a GCM run, we actually compute $\chi_{T_s}^{(1)}(\omega)$ for a specific frequency $\omega = 0$. By using Equations (3-4) and the expression of $\tilde{f}(\omega)$ we obtain:

$$ECS = \frac{2}{\pi}P\int_0^{+\infty}d\omega \frac{Im\left[f_{CO_2}^{2x}\chi_{T_s}^{(1)}(\omega)\right]}{\omega} = \frac{2}{\pi}P\int_0^{+\infty}d\omega\, Re\left[\langle\widetilde{T_s}\rangle_f^{(1)}(\omega)\right] \tag{9}$$

The RRT formalism allows linking the *ECS* to the imaginary part of the susceptibility at all frequencies or, equivalently, to the real part of the response at all frequencies for the imposed step function forcing $f(t)$. Equation (9) clarifies that the values of the response at all frequencies are relevant for determining the long-term response. One may compare the integrand of Equation (9) obtained for two different models in order to test their agreement. In so doing, one would find out which time-scales (and therefore which physical processes) are mostly responsible for possible discrepancies in their *ECSs*. Alternatively, one may find that two models with similar *ECSs* differ substantially regarding their response at different time scales, detecting possible misleading compensating effects.

As mentioned above, the Transient Climate Response (TCR) is defined as the $T_s$ increase at the moment (after about 70 years) [CO₂] has doubled following a 1% per year increase (Otto et al. 2013). In our case $TCR(\tau) = \langle T_s\rangle_{g_\tau}^{(1)}(\tau)$ when the forcing is given by the ramp function $g_\tau(t)$. The forcing is the same used for the scenario of Figure 3, and we obtain $TCR(\tau) = 7.2\,K$. Considering the Fourier representation of $\langle T_s\rangle_{g_\tau}^{(1)}(t)$ evaluated for $t = \tau$ and using Equation (3)

$$\langle T_s\rangle_{g_\tau}^{(1)}(\tau) = \frac{1}{2\pi}\int_{-\infty}^{+\infty}d\omega\,\langle\widetilde{T_s}\rangle_{g_\tau}^{(1)}(\tau)e^{-i\omega\tau} = \frac{1}{2\pi}\int_{-\infty}^{+\infty}d\omega\,\chi_{T_s}^{(1)}(\omega)\tilde{g}_\tau(\omega)e^{-i\omega\tau} \tag{10}$$

The forcing is the sum of the ramp function up to time $\tau$ and the Heaviside function traslated by $\tau$, therefore its Fourier transform is given by $\widetilde{g_\tau}(\omega) = f_{CO_2}^{2x}P\big(\pi\delta(\omega)e^{i\omega\tau} + i\,\text{sinc}(\omega\tau/2)e^{i\omega\tau/2}/\omega\big)$, where $sinc(x) = \sin(x)/x$. Therefore

$$\langle T_s\rangle_{g_\tau}^{(1)}(\tau) = \frac{1}{2}f_{CO_2}^{2x}\chi_\Phi^{(1)}(0) - P\int_{-\infty}^{+\infty}d\omega\,f_{CO_2}^{2x}\chi_\Phi^{(1)}(\omega)\frac{\text{sinc}(\omega\tau/2)e^{-i\omega\tau/2}}{2\pi i\omega} \tag{11}$$

Using the Cauchy formula

$$\frac{1}{2}f_{CO_2}^{2x}\chi_{T_s}^{(1)}(0) = P\int_{-\infty}^{+\infty}d\omega\,\frac{f_{CO_2}^{2x}\chi_{T_s}^{(1)}(\omega)}{2\pi i\omega} \tag{12}$$

we derive

$$\langle T_s \rangle_{g_\tau}^{(1)}(\tau) = f_{CO_2}^{2x}\chi_{T_s}^{(1)}(0) - P\int_{-\infty}^{+\infty}d\omega\, f_{CO_2}^{2x}\chi_{T_s}^{(1)}(\omega)\frac{1+\text{sinc}(\omega\tau/2)e^{-i\omega\tau/2}}{2\pi i\omega} \tag{13}$$

Therefore we can write

$$ECS - TCR(\tau) = INR(\tau)$$
$$= f_{CO_2}^{2x} P\int_{-\infty}^{+\infty}d\omega\, \chi_{T_s}^{(1)}(\omega)\frac{1+\text{sinc}(\omega\tau/2)e^{-i\omega\tau/2}}{2\pi i\omega} \tag{14}$$

The difference between *ECS* and *TCR* is given by a weighted integral of the susceptibility, accounting for the contribution of processes and feedbacks occurring at different time scales. The integral in Equation (14) by KK relations gives $\chi_{T_s}^{(1)}(0)$ in the limit $\tau \to 0$, decreases monotonically with $\tau$, and vanishes in the limit $\tau \to \infty$. $INR(\tau)$ provides a measure of the *inertia* of the system at the timescale $\tau$, due to the overall contribution of the internal physical processes and characteristic time-scales of the relevant climatic sub-systems (Saltmann 2001, Winton et al. 2010). By changing $\tau$, $INR(\tau)$ allows one to deal with different rates of increase of [$CO_2$] (Figure 4). In Figure 4 we also plot $\langle T_s \rangle_f^{(1)}(\tau) - \langle T_s \rangle_{g_\tau}^{(1)}(\tau)$, which instead measures the difference in the transient response at time $\tau$ between the case where the forcing is modulated by $f(t)$ and by $g_\tau(t)$, respectively. This quantity approximately coincides with $INR(\tau)$ for $\tau > 50$ years but has a completely different behavior for small values of $\tau$.

*3.3 Horizons of predictability*

We can quantify the limits to the predictive skills of the theory with a scale analyis. We focus on the high-frequency range. Given an ensemble of N realizations, the error in the estimate of $\widetilde{\langle\Phi\rangle}_f^{(1)}(\omega)$ with respect to the true expectation value can be represented as a random signal $\sigma(\omega)$ such that $|\sigma(\omega)| \approx \alpha N^{-1/2}$, where $\alpha$ is a suitable constant. If the forcing spectrum decays for high frequencies as $|f(\omega)| \sim \beta\omega^{-\nu}$, from Equation (3) we derive that the error on the estimate of the susceptibility is $\left|\delta\left[\chi_\Phi^{(1)}(\omega)\right]\right| \approx \alpha/\beta N^{-1/2}\omega^\nu$. Assuming that the true susceptibility asymptotically decays as $|\chi_\Phi^{(1)}(\omega)| \sim \gamma\omega^{-\kappa}$, the signal to noise ration approaches unity at

$$\omega_c = \left(\frac{\beta\gamma N^{1/2}}{\alpha}\right)^{1/(\nu+\kappa)} \tag{15}$$

For $\omega \gtrsim \omega_c$ the estimate of the susceptibility will be seriously deteriorated by the presence of noise so that we will have no skills in prediction at time scales smaller than $\tau_c = 2\pi/\omega_c$. From our data N=200, we estimate $\alpha \approx 0.2$, $\gamma \approx 0.63$, $\kappa = 1$, while we have by construction $\beta = 1$ and $\nu = 1$. Using Equation (15), we obtain an estimate of $\tau_c \approx 1$ year, which fits with the qualitative information provided by Figure 3. The design of the experiment and the processing of the output involve 3 parameters: the temporal resolution of the response signal $\Delta t$ (tipically a coarse graining of the raw output of the numerical model), the length of the simulations *T*, and the size of the ensemble *N*. They define respectively the high frequency cutoff $\omega_h = \pi/\Delta t$, the low frequency cutoff $\omega_l = 2\pi/T$ (that is also the spectral resolution $\Delta\omega = \omega_l$ ) and the critical value for predictability $\omega_c$. Given the time scales of interest, the appropriate temporal resolution, integration length and ensemble size to have good predictive skills can thus be determined. The quality of prediction differs for different observables and forcings, with *red* observables featuring a slowly decaying susceptibility and forced by *red* modulations being better candidates for a successful prediction.

**4 Summary and discussion**

The applicability of RRT has several important implications regarding climate change science. At a fundamental level, it demonstrates that the problem of climate change is indeed well-defined from a mathematical and a physical point of view, as we can study it in a statistical mechanical framework. Applying the FDT to a non-equilibrum system like the climate can introduce mathematically uncontrollable errors in the estimates of the response or, anyway, requires a very accurate representation of the attractor of the system. While there might be ways to circumvent this problem (Colangeli and Lucarini 2014), we expect to be able to reconstruct the climate response from its natural variability only in special cases. The approach proposed here bypasses some of these mathematical issues by exploiting formal properties of the response and allows for constructing rigorous definitions of climate sensitivity at different time scales through the susceptibility function.

We have provided a framework for relating the difference between transient and equilibrium climate sensitivity to the inertia of the CS, and have shown how these properties depend of the response of the system on all time scales. This partially addresses some issues debated in the literature regarding the specific relevance of the *ECS* (Allen and Frame 2007) and provide viable ways to intercompare GCMs. Deviations from a simple relaxation behaviour have been detected, and point to the complexity of the climate response at multi-annual time-scales. Considering the corresponding spectral features is necessary for having a consistent and integrated picture of the overall climate response at all frequencies. Our approach helps clarifying the limits of simple linear feedbacks approximations.

On the practical side, our results provide a way to perform climate prediction (in an ensemble sense) at all lead-times. We have also shown how to estimate the predictability horizon and assess how it scales with different sizes of the ensemble of simulations. Inaccuracies in representing specific spectral features have serious impacts on our ability to predict climate response on the corresponding time scales, and our findings could help understanding why, e.g., climate response at decadal time scales may be hard to capture. We have highlighted that our ability to predict the response may vary substantially when different observables are considered. It appears that considering only one single $CO_2$ forcing scenario allows for reconstructing accurately the response of the system to other temporal patters of changes in [CO2]. Exploiting the linearity of the response it is possible to compose the effect of multiple forcings (like changes in the solar constant or in other GHGs) by simply adding the linear corresponding Green functions and susceptibilites; the extension to nonlinear cases is more cumbersome but theoretically doable (Lucarini 2009, Lucarini and Colangeli 2012), and basically points to a generalisation of the factor separation technique (Stein and Alpert 1993).

We have limited our analysis to one single observable of primary climatic interest. The analysis of other observables could shed light on the mechanisms determining the response of the CS to $CO_2$ forcing. As an example, the analysis of the response of large-scale meridional gradients of temperature at surface and in the middle troposphere could provide information on changes in the midlatitude circulation. We expect the analysis to work better for globally or regionally integrated, rather than for local, quantities, in terms of signal to noise ratio in the estimate of the susceptibility at high frequencies. In general different observables will show different ranges of linearity. The response of water vapor related observables, in particular, is expected not to scale linearly in the same range as the surface temperature, because of the Clausis-Clapeyron equation (Held and Soden 2006). In this case, if one wants to study large climate shifts the role of the nonlinear terms will likely be relevant. The existence of approximate functional relationship between the susceptibilities of different observables (Lucarini 2009) could set a way for rigorously defining the so-called emergent constraints (Bracegirdle and Stephenson 2013, Cox et al. 2013).

The analysis of fully coupled atmosphere-ocean GCMs is crucial in order to understand the important role of ocean and interactive sea-ice on climate response. Obviously, considering models with interactive ocean introduces longer time scales and makes the analysis technically more difficult. For a prediction in an ensemble sense, we need to be able to account for all time scales. This implies the use of very long integrations and a large number of ensemble members. A tentative intercomparison analysis on several GCMs could be performed using CMIP5 data, even if this goal seems to be out of reach given the limited length of the hosted runs. This hints at the need of including in model data repositories longer runs to properly assess the effect of the climate change scenarios on the state of the system (Lucarini and Ragone 2011). As a first (still challenging but currently more doable) step in the direction of testing RRT in state of the art climate models, we are currently working on the large ensemble of simulations made available in the climateprediction.net project.

RRT provides a well defined theoretical framework and tools that allows to diagnose rigorously discrepancies in the properties of the frequency dependent response of different models and to guide the design of the climate change experiments. On one side, this could support strategies for future GCM development aimed at improving the persisting deficiencies in the model performances highlighted in the last IPCC report (IPCC 2013) by exploiting rigorous analysis and intercomparison procedures. On the other side, the vast number of diverse scenarios considered nowdays in the IPCC protocol could be substituted by a more focused effort on few selected forcing experiments making use of ensemble methods. Or, conversely, given the present palette of forcing scenarios, one could fill in the gaps and create projections for combinations of scenarios without resorting to additional simulations. This is promising for applications, considering that simulations for climate change assessment are extremely expensive and, at the same time, it is necessary to inform policymakers on a wide range of climate change scenarios.

**Acknowledgments**

The authors wish to thank C. Franzke and G. Gallavotti for commenting on an earlier version of the manuscript. F.R. wish to thank T. Bodai and S. Schubert for useful discussions. F.R. and V.L. acknowledge fundings from the Cluster of Excellence for Integrated Climate Science (CLISAP) and from the European Research Council under the European Community's Seventh Framework Programme (FP7/2007-2013)/ERC Grant agreement No. 257106. The authors acknowledge the Newton Institute for Mathematical Sciences (Cambridge, UK), hosting the 2013 programme "Mathematics for the Fluid Earth" during which part of this work was discussed.

**Figures**

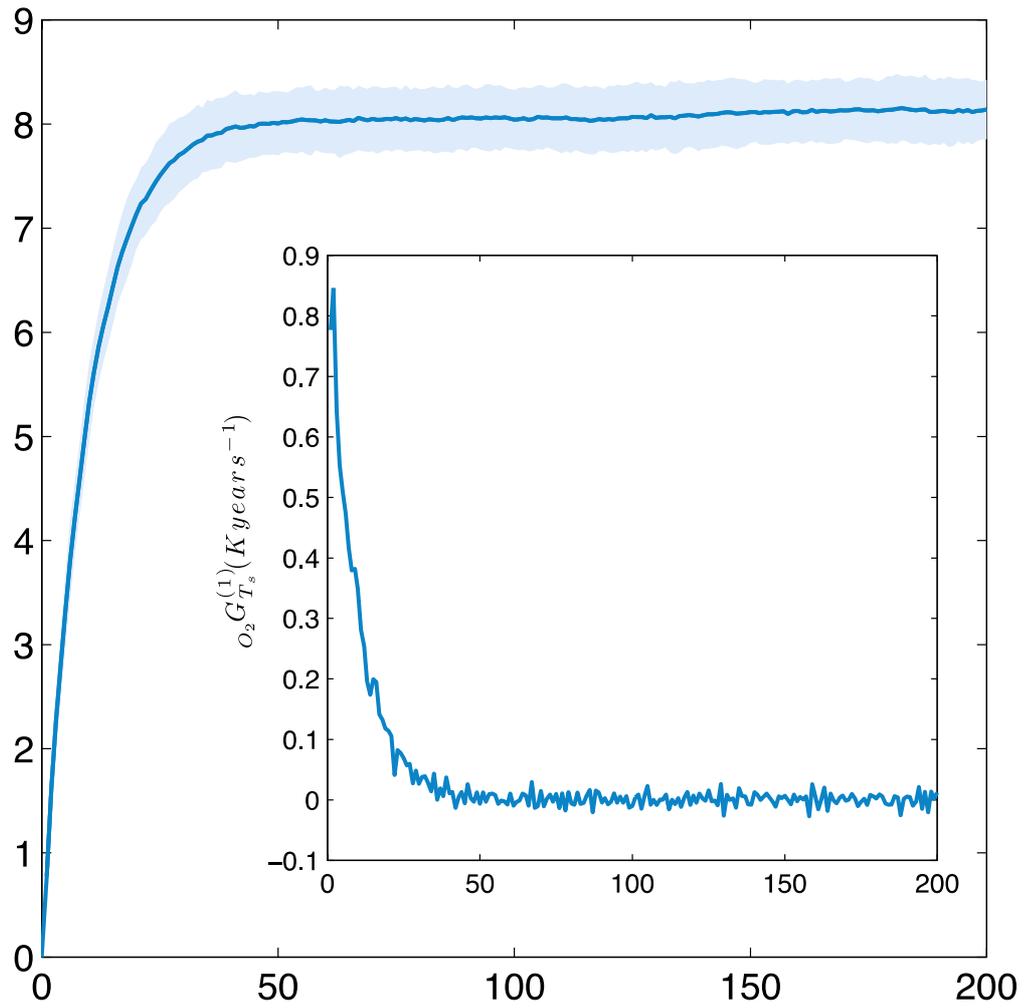

**Figure 1** Ensemble average of increase of global surface temperature after an instantaneous increase of the [CO2] by a factor 2. The upper and lower limit of the bands are computed as two standard deviations of the ensemble distribution. Insert: Green function of the global surface temperature.

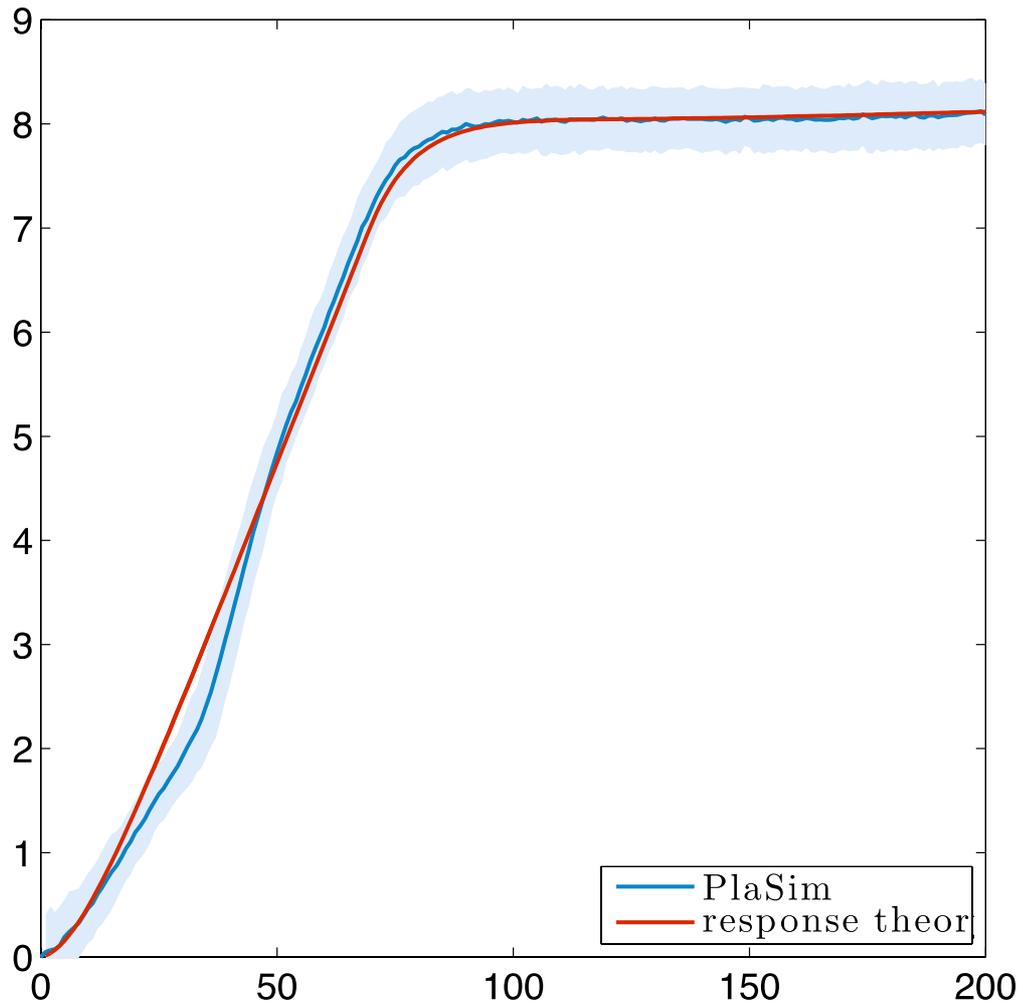

**Figure 2** Comparison between GCM simulation (blue) and response theory prediction (red) for 1% per year increase of the CO2 concentratrion. The upper and lower limit of the bands are computed as two standard deviations of the ensemble distribution. The agreement between the GCM run and the prediction via linear response theory is remarkable, with only a slight discrepancy during the transient, most probably connected to the activation of the ice-albedo feedback.

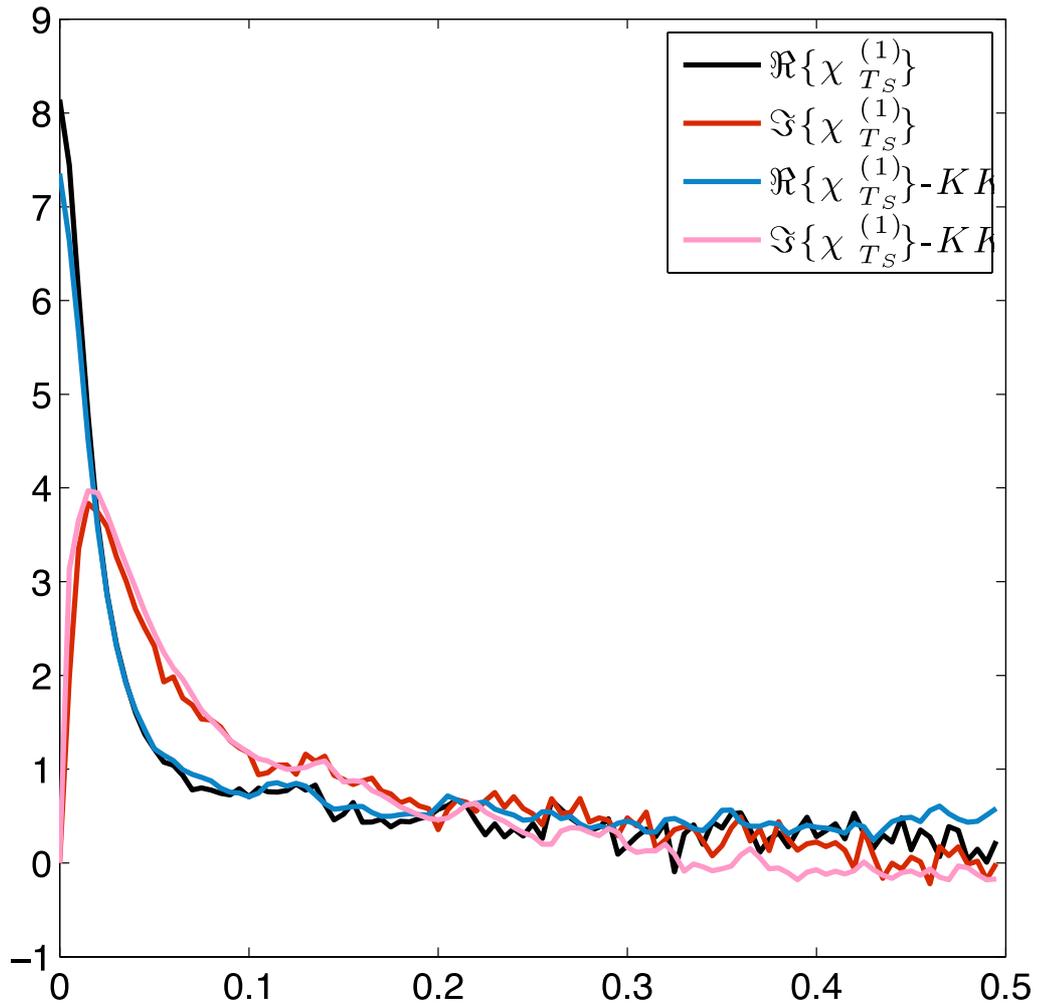

**Figure 3** Real (black) and imaginary (red) parts of the susceptibility computed as the Fourier transform of the Green function of Figure 1. Real (blue) and imaginary (pink) parts of the susceptibilty obtained via singly subtractive KK relations.

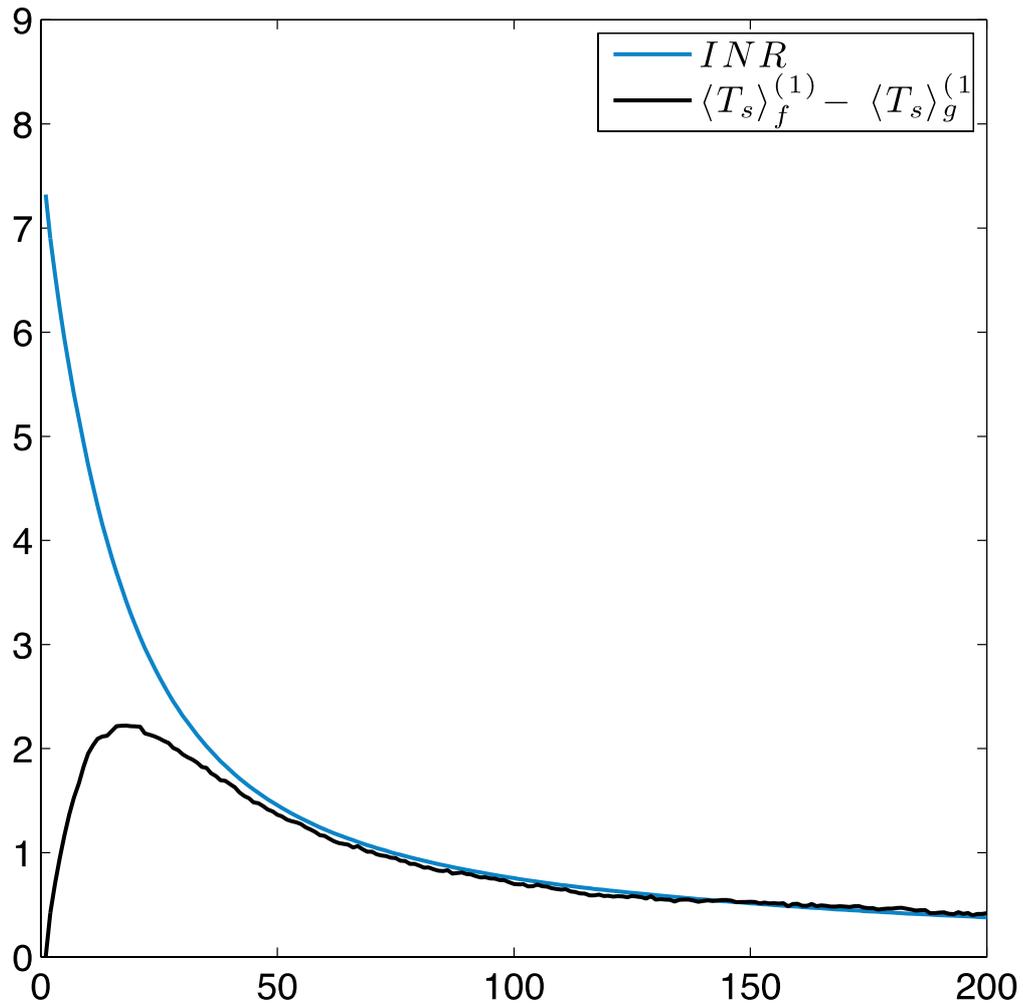

**Figure 4** Measuring the relaxation of the system. Inertia of the system at time scale $\tau$ (blue) and difference at lead time $\tau$ between the response at the system to a step-like and ramp (of time scale $\tau$) forcings (black).